# Long Sync Word Frame Synchronization for Future Wireless Networks

***Dimitris Nikolaidis***

## Abstract

Frame synchronization is the act of accurately detecting frames in an incoming transmission and extracting their payload. It is especially important in environments such as wireless channels where signals are significantly distorted. Digital correlation is the simplest form of frame synchronization where XNOR gates and summations are used to perform correlation and detect binary sync words attached to each frame. In this paper we demonstrate how digital correlation with long syncwords in the context of a standard modem (modulator/demodulator) chain solves the problem of frame synchronization under significant signal distortions in a practically implementable manner with minimal extra resources. The scheme consists of two parts, the modem chain and the frame synchronization architecture. The modem chain is built inside GNU radio software tool and the architecture inside VIVADO. Data transmission and reception in the Rayleigh NLOS channel is performed by the modem and frame synchronization by the architecture with data produced by GNU radio via a verilog testbench. Results indicate high synchronization accuracy with distortion levels much higher than what is reported by other methods.

## Introduction

Frame synchronization refers to the accurate detection of the starting bit of the frame by the receiver. It is particularly important in scenarios where there is intense distortion imposed on the transmitted signal. One such field is wireless communications where wireless channels are characterized by significant bit error rate even with moderate levels of noise. There are various proposed schemes that perform frame synchronization such as machine learning aided synchronization[1][2], joint error correction and frame synchronization [3][4] or maximum likelihood metrics[5][6] but the traditional and simplest method to perform frame synchronization is to apply digital correlation on a sync word attached to the frame and use the correlation peak to ascertain the position of the starting bit [7][8].

Digital correlation of two binary vectors entails the appliance of XNOR gates to bits in respective positions and the summation of the products of the gates. This means that the value of the summation indicates how many similar bits the two vectors have. To detect the frames and extract their payloads, a predefined syncword is attached to every frame. The received bitstream is then correlated with the predefined syncword by utilizing the sliding/moving window technique to detect the correlation peaks and find the frames.

In this paper we demonstrate how digital correlation with very long binary sync words solves the problem of frame synchronization in wireless links more successfully than other state of the art methods currently available. This is achieved by testing a modified version of the architecture described in [9] in the context of a basic wireless modem (modulator/demodulator) process chain [10] under unfavorable conditions. The architecture utilizes a long sync word correlation scheme that can detect long sync words accurately and extract the payload of the frame while consuming minimal extra hardware resources at very high line rates. The modem chain together with the integrated frame synchronization architecture constitutes a complete digital transmitter/receiver chain with capabilities aligned with future wireless network requirements.

Testing is performed with the aid of GNU radio [11], a reliable open source python-based software tool used to simulate communication signal processing chains. More specifically, a bitstream consisting of multiple frames is first created in MATLAB and recorded to a binary file. The binary file is then read by the modem process chain built inside GNU radio and transmitted over Rayleigh NLOS channel. It is then demodulated and recorded in binary form to another binary file. This file is fed to the architecture via a VIVADO verilog testbench. The architecture runs on the bitstream and detects and extracts what data it can. The data is presented as output to and recorded by the verilog testbench in a final binary file. The extracted data in the final file is then compared with the original bitstream to calculate the percentage of the missing frames.

When compared with state of the art methods the proposed scheme offers much higher accuracy with very limited extra hardware and power requirements when tested in very unfavorable conditions (low SNR Rayleigh NLOS). It can also potentially support high line rates due to its minimalist symbol sync stage and efficient frame synchronization architecture. It is superior in efficiency to both standard frame synchronization schemes such as maximum likelihood and joint error correction and frame synchronization as well as novel methods involving frame synchronization with AI. Moreover its effectiveness persists in distortion heavy environments (Rayleigh channel model) and across multiple link specifications (syncword size). It can also be used to provide data security by masking the frames with impossible to detect long syncwords. It is modulation independent and can be used in conjunction with techniques like adaptive modulation [12].

## Scheme

The paper showcases how digital correlation with very long binary syncwords can be effectively used to enable wireless transmission under extremely unfavorable channel conditions not possible with current state of the art hardware. The bitstream to be transmitted is read from a binary source file and the demodulated received bitstream is recorded to a different binary sink file. The bitstream is divided in data frames and each data frame consists of payload ($n$ bits) and attached syncword ($k$ bits) where $n>>k$, totaling $n+k$ bits per frame. Bitstream generation is achieved with MATLAB. Syncwords are long ($k>200$) and are detected through digital correlation after reception. They are generated randomly in the sense that they are a product of a random binary sequence generator before the start of the transmission and they are known to both transmitter and receiver. The same syncword applies for a certain number of frames (for example $10^5$ frames). Implementation of the scheme is achieved by transmitting the data using GNU radio wireless channel simulation tools, recording the transmitted bitstream on a separate file and then inserting it as input to the architecture using Xilinx VIVADO verilog testbench.

A basic modem chain transmission consists of framing and modulation encoding. Framing is where raw digital data is encoded with the selected error correction scheme and organized into different frames for transmission. Modulation encoding transforms the binary data of the frames to constellation points which are then transmitted as waveforms [13]. There is a lot of variation in the methods used in both stages to achieve this result but in general terms this is the process of basic wireless transmission. For the proposed scheme, data read from the file is ready for transmission (separated in frames with sync words attached and assumed to be encoded) so modulation encoding is the only stage present in the modem chain. Moreover the source of the data is a binary file read with GNU radio block file source. After modulation the signals passes through the channel (fading model for Rayleigh fading and AWGN model for white noise addition) that simulates transmission and is received by the receiver.

The receiver side is more complicated than the transmitter. The first part is symbol synchronization and happens exactly after the information signal has been received. There are multiple algorithms used for symbol synchronization[14] but in our case we chose maximum likelihood[15]. The signal is filtered and a symbol synchronization algorithm together with the costas loop scheme[16] is applied to ascertain the symbol period and synchronize the frequency of the signal under the presence of noise, frequency and timing shift. Once the symbol period has been estimated the symbol decoder/ demodulator circuit uses the down sampled available data from frame sync and costas loop to correctly decode the symbols into bits and reproduce the demodulated bitstream as it was transmitted by the transmitting system before modulation.

The complete diagram of the modem is presented on figure 1. It is a slightly modified version of the tested GNU radio QPSK modem example found in the tutorial section of GNU radio webpage [11]. Simulations that test the scheme's functionality were run on this model. It is important to note that the symbol synchronization algorithm was changed to sign of max likelihood as it is the more implementation friendly version of max likelihood.

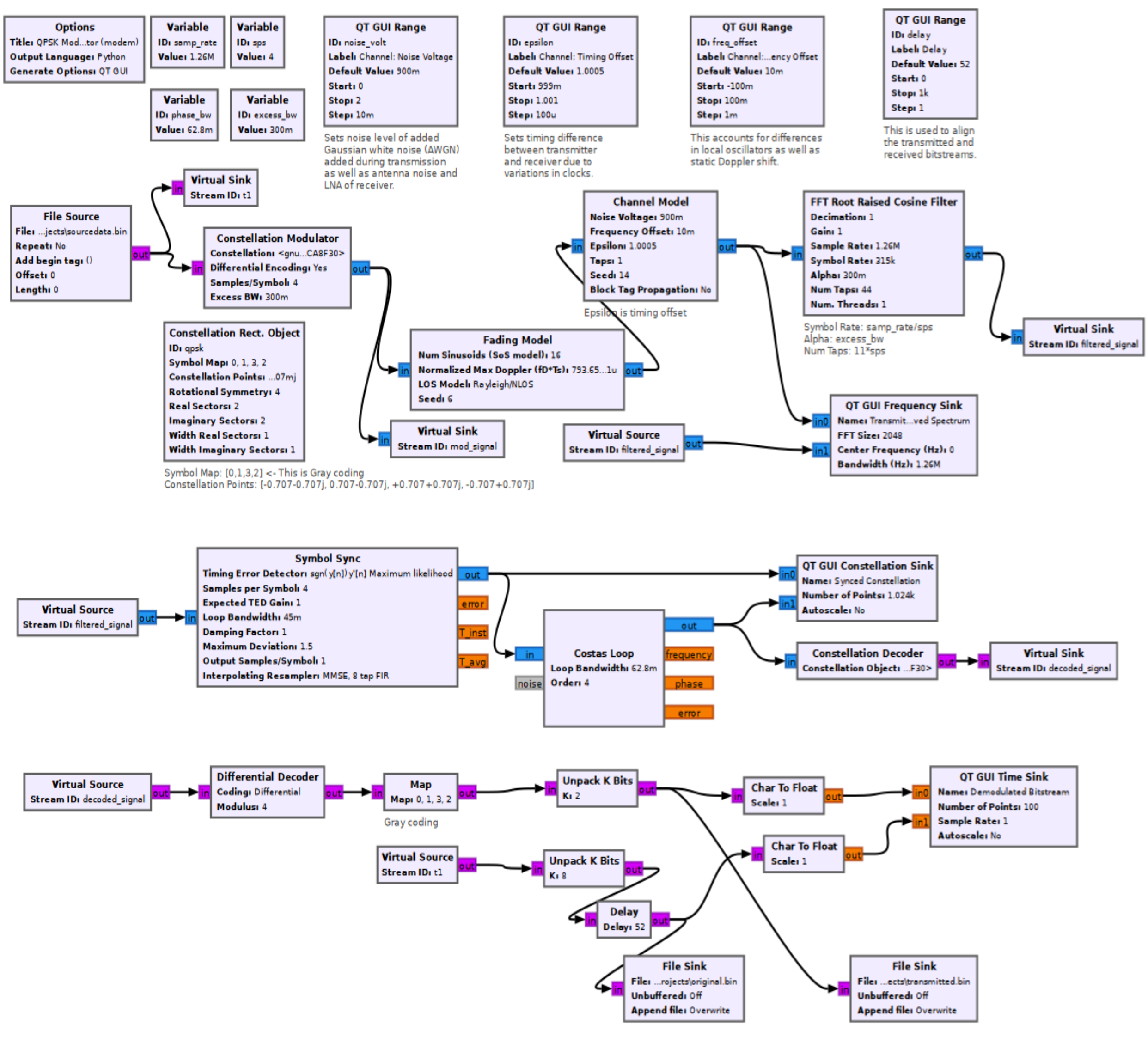

Figure 1: Complete flow diagram of the modified QPSK modem chain. Channel is simulated with fading and AWGN blocks for Rayleigh fading and white noise addition respectively. Source and destination of data are binary files.

Raw random data comes from a binary file (datasource.bin in file source block) which is assumed to be transmission-ready data. After modulation the signals passes though the simulated communication channel. Block virtual sink connects the output of the filter block to the other part of the receiver through virtual source. Blocks symbol sync and costas loop together constitute the timing synchronization stage and block constellation decoder transforms the time synchronized stream to a demodulated binary bit stream. The list of parameters were defined by the original example and include frequency offset, timing offset, noise voltage, excess bandwidth, sps (samples per second), sample rate, delay. Three of them, frequency offset, timing offset and noise voltage are presented as separate blocks inside the app connected to a controllable GUI. The rest are presented as variable blocks.

After demodulation, the received bitstream is recorded in file transmitted.bin and fed to the architecture through a verilog testbench. The original transmission is also recorded in file original.bin.

The frame synchronization architecture used is a modified, much more hardware efficient version of the one in [9], based on the modification made in [17]. It scans the datastream for the predefined sync word through digital correlation. Upon detection, the architecture extracts the frame payload and the verilog testbench records the extracted data to a final binary file. The original bitstream is then compared with the architecture output data recorded by the testbench in MATLAB to calculate the number of missed frames. The entire workflow of the process can be seen in Figure 2.

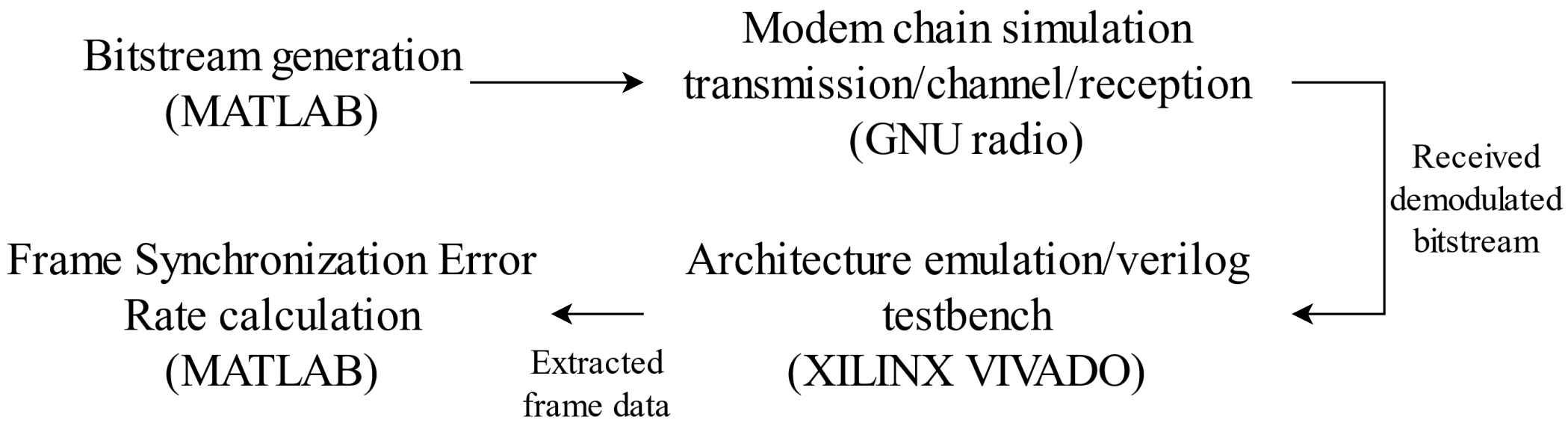

Figure 2: Input bitstream is generated in MATLAB as a binary file. It is read by GNU radio and its transmission over a wireless channel is simulated in the GNU companion graphical app. The demodulated transmitted binary data is recorded to a new file which is then fed to the architecture through a verilog testbench. The architecture runs on the data, detects the frames, extracts the payload and outputs it so that the verilog testbench can record it to a final binary file. The final binary file is compared with the original bitstream through MATLAB to calculate the percentage of missing frames.

Digital correlation in the context of two binary vectors of equal length is performed by applying XNOR gates to bits in respective positions and adding the products of the gates. The XNOR gate is also called equivalency gate because its output is the digital 1 only when both inputs are the same. This results in a value that indicates the number of similar bits between the two vectors. To find the syncword inside the bitstream, the architecture uses the sliding window operation and slides the predefined syncword across all possible positions inside the stream, performs digital correlation and determines whether the correlation value (number of similar bits) is the possible start of the frame or not depending on a set threshold. If the value of the correlation sum is above the threshold at a specific bit position inside the bitstream it serves as a sign that a syncword and frame might be present and the payload capture process begins.

To execute the sliding window operation the architecture combines a register with parallel adder trees. The register acts as a buffer for the bitstream and the parallel adder trees perform the correlation operation for every possible position of the syncword in parallel. Adder trees are circuits who provide the summation of a set of numbers and their structure is similar to binary trees. In each level numbers are added by two until the final summation is calculated after $\log_2$ steps. Due to their flexible form they can be implemented efficiently by utilizing techniques like pipelining. If the highest value of all tree outputs surpasses the predefined threshold the architecture assumes a syncword has been detected at that position and begins the payload capture process. Payload capture is achieved with a delay register. Selection of the highest value was implemented as a comparator tree. A module with the same topology as the adder tree but with comparators instead of adders. Figure 3 presents the frame detection circuit abstractly.

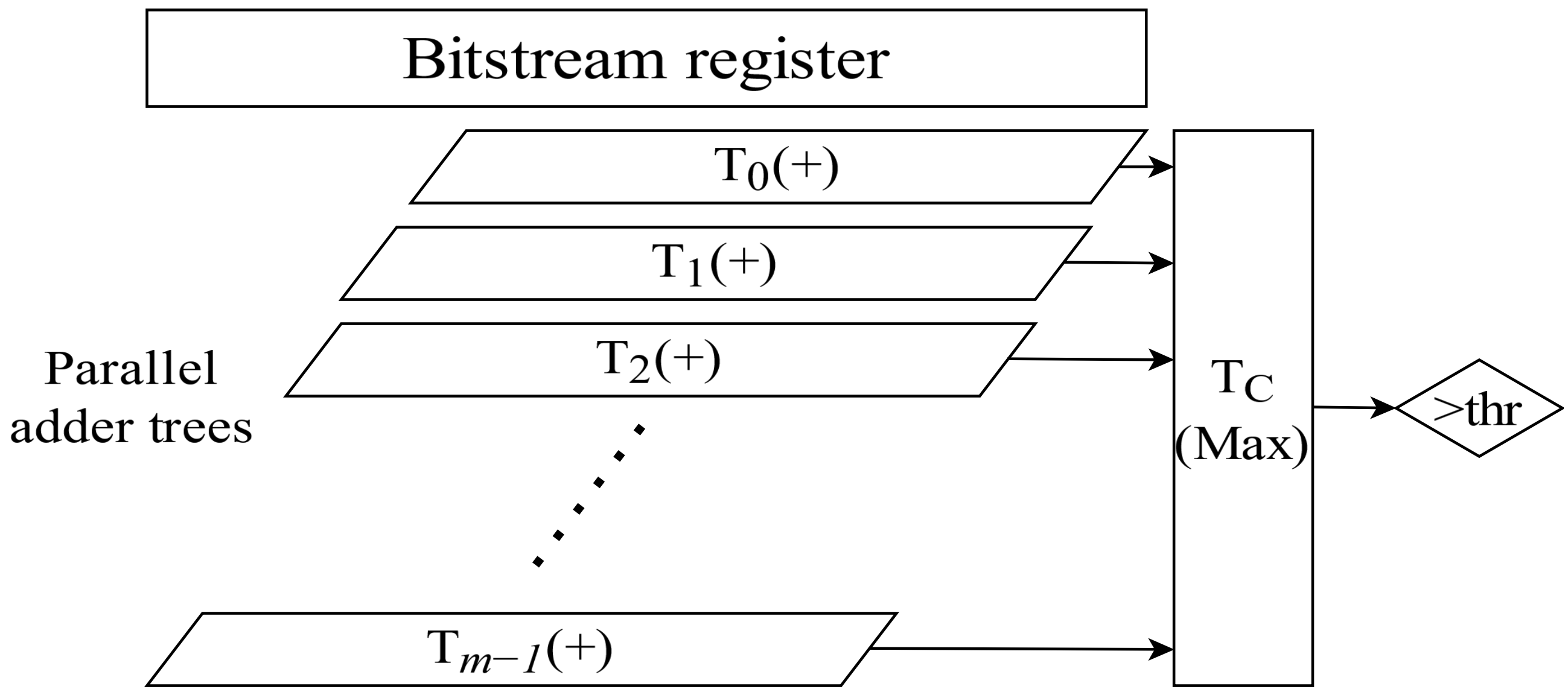

Figure 3: Adder trees ($T_i$) perform correlation in consecutive positions in parallel, implementing the sliding window scheme on the bitstream register. The maximum correlation value is selected with the comparator tree ($T_c$) and compared with the threshold. The length of the sliding window is *m* bits.

Even though digital correlation is not a novel method and has been used before, there is a difference between simply using digital correlation and using digital correlation with very long sync words. For short sync words (<50 bits) other schemes like joint error correction and frame synchronization and maximum likelihood produce better results. It is only when the sync word length becomes very large that digital correlation becomes substantially more accurate.

This can also be shown mathematically. The syncword detection process is the equivalent of an error correction code with a code rate of $1/k$. The $k$ bits of the syncword are translated to 1 bit, which can take two values. 0 if the correlation value is below the threshold and there is no detection and 1 if it surpasses the value and there is detection. According to Shannon Theorem if $k$ is very large, error less transmission is possible even for very high noise. In the context of the code and architecture, error less transmission is the correct detection of the first bit of the syncword and frame through the threshold surpassing correlation peak.

The architecture in [9] is the only method that can perform this task while being practically implementable for very large sync word size with minimal hardware consumption. It is constructed with only basic hardware elements that are organized in an efficient structure with optimal logical complexity and thus delay, allowing for the utilization of techniques such as parallelization and pipeline that further increase its efficiency in terms of throughput. The modem chain together with the integrated architecture constitutes a complete functional digital transmitter/receiver which surpasses state of the art wireless transmission hardware in accuracy and rivals it in terms of cost and throughput. It is also important to note that the architecture itself is modulation independent. In this particular example a QPSK modem was used to test its accuracy but since it acts upon the binary demodulated bitstream it can be used with techniques like adaptive modulation [12] where the modulation scheme dynamically changes. Moreover, the random nature of the long syncwords makes them impossible to detect for an outsider concealing the transmission.

## Results

The architecture was tested under the Rayleigh NLOS channel model with the respective specialized function blocks in GNU radio. It was simulated for two syncword sizes, 300 and 500 with thresholds 210 and 350 respectively, across a range of noise voltages values. Timing offset and frequency offset were kept to their default value given by the original example for all simulations. The input bitstream contained $2 \times 10^4$ frames in total. Noise voltage is what was changed primarily during simulations along with the seeds of the channel model blocks to ensure complete randomness. The range of noise voltage values was set to [0.4-1.1] with step equal to 0.1.

Implementation data (hardware resources/line rate/platform of implementation) for the architecture for the specific syncword sizes is also provided to give a general estimate of the schemes required resources. LUTs and Ffs are FPGA resources used to implement circuits. The platform of implementation is board Xilinx KCU116. Line rates above Gigabit are achieved with less than 10% of resources of a mid-range FPGA.

| framesize | LUT% | FF% | Line Rate | Power | Power/Rate |
|---|---|---|---|---|---|
| 300 | 9139(4.21%) | 14429(3.33%) | 3.75 Gbps | 0.959 W | 0.26 |
| 500 | 20237(9,33%) | 31009(7.15%) | 5 Gbps | 1.495 W | 0.3 |

Figure 4 presents the frame synchronization error rate (FSER) I.e the percentage of lost frames in NLOS Rayleigh channel for the two sync word sizes for the equivalent to noise voltage SNR range in dB. For the Rayleigh channel, 16 sinusoids were used. The normalized doppler frequency was set to $10^{-3}$.

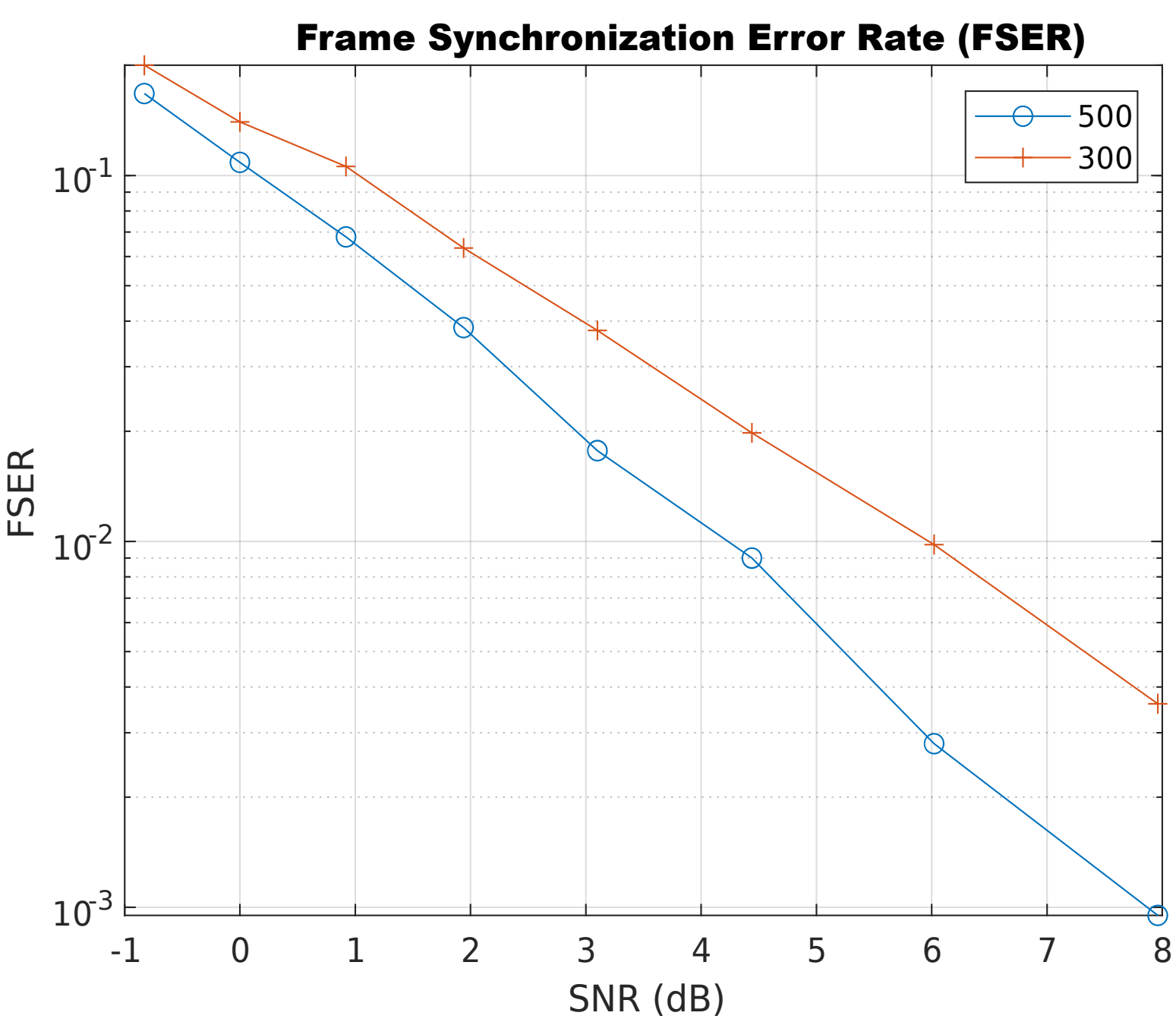

Figure 4: FSER for Rayleigh channel (16 sinusoids, $10^{-3}$ normalized Doppler frequency)

As expected, FSER degrades with higher SNR and longer sync words have higher synchronization accuracy when the threshold is proportional. The 500 bits version has higher accuracy than the 300 version with different absolute threshold (350 to 210) since 350/500=210/300=0.7.

Lowering the threshold of the 300 version may result in different results however there is a limit. It is recommended that only long syncwords are used with the scheme (>250 bits) to avoid false detections. This phenomenon occurs when patterns with high correlation values appear naturally inside random data and trigger the threshold detection mechanism. If the syncword is short this happens often however if its long then the probability is practically zero. Also, transmission fails near -1 dB not because of the architecture but because the symbol sync block fails to perform proper timing synchronization resulting in demodulation failure. It is possible to push the architecture even further by implementing a more resilient symbol sync algorithm for low SNR than what is available in GNU radio.

The accuracy of the method either surpasses or is similar to other schemes even though it was tested under much harsher channel conditions. [18] and [19] are comprehensive publications that offer comparisons between the classical maximum likelihood correlation method with modern error correction and frame synchronization schemes based on LDPC codes. The best result is a loss between $10^{-1}$, $10^{-2}$ of frames around 0 dB in AWGN channel. In [5] and [21] max likelihood methods are used to achieve frame synchronization. The reported accuracy at SNR=0dB is around 0.95 (0.05 FSER). In general across different methods synchronization rates seem to cluster between $10^{-1}$ and $10^{-2}$ in the pure AWGN channel. Considering that the BER in AWGN is magnitudes worse than Rayleigh this accuracy is much worse that our reported result. Frame synchronization in Rayleigh fading is rarely seen in literature due to the time varying nature of the channel which makes consistent detection of frames difficult. [21] is one of the few publications that examine frame synchronization in flat fading Rayleigh channel. Even when accounting for the 3dB difference between coherent and incoherent modulations [22] the reported frame detection rate is around 0,5 at 3 dB SNR, much lower than the proposed scheme.

In terms of practical implementation the entire modem chain and synchronization architecture can be easily implemented. Modulation and Demodulation are standard circuits and there is a lot of available implementation methods. Symbol synchronization requires more resources however both quantized maximum likelihood symbol synchronization [23] and costas loop implementations [24] are available in literature. Addition of error correction encoders/decoders can also be considered when attempting to implement a commercial modem.

## Conclusion

This paper demonstrates the merits of long binary syncword correlation frame synchronization in digital links. The entire scheme can be implemented easily by integrating the frame synchronization architecture into a standard modem (modulator/demodulator) chain. Due to the architecture's minimalist nature, this can be achieved with limited extra resources and support very high line rates. Its long sync word correlation accuracy is highly reliable and can provide stable frame synchronization under levels of noise unmanageable for state of the art methods. It is modulation independent as it acts only on the demodulated digital bitstream so it can be implemented with techniques like adaptive modulation. Moreover it can be used as a security mechanism to conceal the start of the frame due to the random nature and length of the sync word. It fulfills all requirements to be considered a viable option for future generation wireless networks.